\documentclass[aps,prl,reprint,groupedaddress,showpacs,showkeys,amsmath,amssymb]{revtex4-2}

\usepackage{graphicx}
\usepackage{dcolumn}
\usepackage{bm}
\usepackage{xcolor}
\usepackage{physics}

\begin{document}
	
	\title{Theory of Multiple Scattering Enhanced Single Particle Plasmonic Sensing }%
	\author{Joel Berk}
	\author{Matthew R. Foreman}%
	\email{matthew.foreman@imperial.ac.uk}
	\affiliation{%
		Blackett Laboratory, Imperial College London, Prince Consort Road, London, United Kingdom SW7
		2BW}%

	\date{\today}
	
	\begin{abstract}
Methods to increase the light scattered from small particles can help improve the sensitivity of many sensing techniques. Here, we investigate the role multiple scattering  plays in perturbing the scattered signal when a particle is added to a random scattering environment. Three enhancement factors, parametrising the effect of different classes of multiple scattering trajectories on the field perturbation, are introduced and their mean amplitudes explored numerically in the context of surface plasmon polariton scattering. We demonstrate that there exists an optimum scatterer density at which the sensitivity enhancement is maximised, with factors on the order of $10^2$ achievable. Dependence of the enhancement factors on scatterer properties are also studied.
	\end{abstract}
	
	\maketitle
	High sensitivity and label-free optical measurements play a critical role in applications including clinical diagnostics, environmental monitoring and detection of single nanoparticles \cite{Fan2008,Zhu2014Light-ScatteringNanoparticles}. Detection strategies employing light scattered from analyte particles, such as dynamic light scattering and interferometric scattering microscopy, have proven highly successful \cite{Stetefeld2016DynamicSciences,Taylor2019InterferometricScattering,Yang2018InterferometricExosomes,Zhang2020PlasmonicKinetics} {with detection of discrete binding events of biomolecules such as proteins \cite{Ferreira2009prt,Liu2009prt}, virions \cite{LeeNov2019}, DNA \cite{Zhao2003DNA} and enzymes \cite{LiEnzyme}, representing one of the ultimate goals in the field}. Performance of such systems can, however, significantly degrade in the presence of additional secondary or multiple scattering from the local environment \cite{Badon2017MultipleMicroscopy,Yoon2020DeepMedia,Bar-Ziv1997LocalizedNanoscale}.  In many systems of experimental interest, for example colloids or biological tissue, multiple scattering is unavoidable and must thus be accounted for in order to probe them accurately \cite{DWS88,Mosk2012ControllingMedia}. Multiple scattering effects however also afford a number of practical gains. For example, the inherent angular spread caused by scattering allows the diffraction limit to be overcome \cite{Choi2011}, whilst random optical speckle patterns have been shown to possess sensitivity to the properties of a single particle \cite{Berkovits1991,Nieuwenhuizen1993} in turn enabling their localisation \cite{Berkovits1990,denOuter:93}. Such potential advantages mean that engineering the photonic scattering environment in order to promote multiple scattering is frequently investigated.  Generation of small regions in which the electric field intensity is much larger than the surrounding region, using for example, metallic nanoparticles near metal interfaces \cite{Huang2017Nanoparticle-on-mirrorEnhancement, Baumberg2019} or rough metal surfaces \cite{Cang2011ProbingImaging}, is a common example. Analyte particles in such `hotspots' in turn scatter more light 
thereby endowing sensors with a greater sensitivity \cite{Alonso-Gonzalez2012ResolvingSpots}. Similar hotspot mechanisms have been studied in the context of enhanced fluorescence and Raman scattering \cite{Cang2011ProbingImaging,Liu2014Three-DimensionalMatrix,Itoh2017One-dimensionalScattering}. Carefully designed nano-structured substrates have also received significant attention  \cite{Stewart2008, Konopsky2018PhotonicPeak}, whereby coupling of different nanostructures can augment any perturbation upon addition of an analyte particle. Randomly distributed nanostructures are also known to give rise to a rich set of multiple scattering phenomena not seen in deterministic structures, such as Anderson localisation and long range correlations \cite{Segev2013,Maystre1994,Boguslawski2017,Berkovits1994}, which can aid single particle detection. In combination with the less stringent fabrication requirements, random sensors therefore represent a particularly promising platform for enhanced particle sensing.

In this letter we consider the origin and magnitude of differing mechanisms which can enhance single particle sensing in random multiple scattering environments. Three classes of scattering trajectory are analytically identified corresponding to coupling between different scatterers, generation of localised hotspots and scattering induced self-interactions. We show, through numerical modelling of a random nanostructured plasmonic substrate, that competition between different multiple scattering effects, namely dipolar coupling and localisation, provides opportunities to optimise achievable enhancements through variation of the average scatterer density and polarizability. Insights gained in this work can hence guide future design of optimal scattering based single particle detectors in turn facilitating for instance non-equilibrium biological studies \cite{nonequil} or study of molecular machines \cite{molmac}.

In order to study a disordered scattering environment, we use a coupled dipole formalism, valid for scattering from small scatterers in which the dipole mode is dominant \cite{Novotny1997InterferencePlasmons,Chaumet2005EfficientMethod,Sndergaard2003VectorialInteractions}. Typically, the dipole approximation is valid for sub-wavelength size scatterers and when the field within the scatterer is approximately homogeneous \cite{Novotny2006}.   Initially we consider a system of $N$ dipole scatterers centred at $\bm{r}_i$ ($i=1,2,\ldots,N$). When illuminated with a monochromatic electric field $\bm{E}_0(\bm{r})$ of frequency $\omega$, the total field $\bm{E}(\bm{r})$ at a position $\bm{r}$ outside the volume of the scatterers is
\begin{equation}\label{eq:coup_dip}
	\bm{E}(\bm{r})=\bm{E}_0(\bm{r})+\frac{k_0^2}{\varepsilon_0}\sum_{j=1}^N G(\bm{r},\bm{r}_j)  \bm{p}_j,
\end{equation}
where $k_0=\omega/c$ is the free-space wavenumber, $c$ is the speed of light in a vacuum, $\varepsilon_0$ is the vacuum permittivity and $G(\bm{r},\bm{r}')$ is the Green's function defined with respect to the background dielectric function $\varepsilon(\bm{r})$, i.e. excluding the $N$ scatterers. Notably, we allow the dielectric function to vary spatially such that our description is applicable to substrate based setups. The dipole moment of the $j$th scatterer is given by $\bm{p}_j=\alpha_j\bm{E}_\text{exc}(\bm{r}_j)$ where $\alpha_j$ is the dressed polarizability including any potential self-interactions (e.g. due to reflections from a substrate) and $\bm{E}_\text{exc}(\bm{r}_j)=\bm{E}_0(\bm{r}_j)+\sum_{i\neq j}G(\bm{r}_j,\bm{r}_i)  \bm{p}_i$ is the exciting field incident on the $j$th dipole, consisting of the incident field and the field from all other dipoles \cite{Novotny1997InterferencePlasmons,LakhtakiaCDA}. In general, $\alpha_j$ is a tensor, however reduces to a scalar for isotropic scattering, e.g. spherical scatterers in a homogeneous environment. The set of equations $\bm{p}_j=\alpha_j\bm{E}_\text{exc}(\bm{r}_j)$ can be expressed as the set of linear equations:
\begin{equation}\label{eq:coup_dip_M}
	\sum_{j=1}^NM_{ij}\bm{p}_j=\bm{p}_{0,i},\quad ~i=1,2,\ldots,N
\end{equation}
where $\bm{p}_{0,i}=\alpha_i\bm{E}_0(\bm{r}_i)$ is the dipole moment of the $i$th scatterer induced solely by the incident field,
\begin{equation}\label{eq:M_def}
	M_{ij}=\begin{cases}
		I_3\quad &i=j\\
		-\frac{k_0^2}{\varepsilon_0}\alpha_iG_{ij} \quad &i\neq j,
	\end{cases}
\end{equation}
$I_3$ is the $3\times 3$ identity matrix and for convenience we let $G_{ij}=G(\bm{r}_i,\bm{r}_j)$. Formally, the dipole moments are given by $\bm{p}_i=\sum_{j=1}^NM_{ij}^{-1}\bm{p}_{0,j}$. Note $M_{ij}^{-1}$ denotes the $(i,j)$th $3\times3$ block (i.e. rows $3i-2$ to $3i$ and columns $3j-2$ to $3j$) of the inverse of the entire $3N\times3N$ coupling matrix, as opposed to $(M_{ij})^{-1}$, the inverse of the $3\times3$ matrix $M_{ij}$. In the single scattering regime, interactions between different scatterers are negligible such that $M_{ij}=M_{ij}^{-1}=I_3\delta_{ij}$ and $\bm{p}_{i}=\bm{p}_{0,i}$.

Introduction of an additional scatterer, namely the analyte particle, with polarizability $\alpha_{N+1}$ at position $\bm{r}_{N+1}$, to the disordered system produces an associated change in the scattered field, $\delta\bm{E}(\bm{r})$, given by
\begin{equation}\label{eq:delta_E}
	\delta\bm{E}(\bm{r})=\frac{k_0^2}{\varepsilon_0}{G}(\bm{r},\bm{r}_{N+1}) \bm{p}_{N+1}+\frac{k_0^2}{\varepsilon_0}\sum_{j=1}^N {G}(\bm{r},\bm{r}_j)  \delta \bm{p}_j.
\end{equation}
The first term in Eq.~\eqref{eq:delta_E} corresponds to the additional dipole field originating from the analyte particle, whilst the second term is the change arising due to the perturbations to the original $N$ dipole moments $\delta\bm{p}_j$. Analogous expressions for the perturbed field have been derived previously within a scalar model \cite{Huang2004ATheory} in terms of the determinant of the coupling matrix, however, the vectorial form in Eq.~\eqref{eq:delta_E} is more appropriate for electromagnetic problems. Within the single scattering approximation, there is no coupling between dipoles whereby $\delta\bm{p} = \bm{0}$ and the perturbation to the scattered field $\delta\bm{E}_{ss}$ reduces to 
\begin{equation}\label{eq:delta_E_ss}
	\delta\bm{E}_{ss}(\bm{r})=\frac{k_0^2}{\varepsilon_0}{G}(\bm{r},\bm{r}_{N+1}) \bm{p}_{0,N+1}.
\end{equation}
In the full multiple scattering case the perturbation $\delta\bm{E}$ can be expressed in the same form as Eq.~\eqref{eq:delta_E_ss} albeit with a modified dipole moment $\bm{p}_{0,N+1}\to\gamma_1\gamma_2\gamma_3 \bm{p}_{0,N+1}$ (see Ref.~\citenum{Berk2021} for a full derivation) where $\gamma_1$, $\gamma_2$ and $\gamma_3$ are enhancement factors given by
\begin{align} 
	\gamma_1&=I_3+\frac{k_0^2}{\varepsilon_0}G(\bm{r},\bm{r}_{N+1})^{-1}\nonumber\\
	& \quad\quad\quad\quad\times\sum_{i,j=1}^N {G}(\bm{r},\bm{r}_i){M}_{ij}^{-1} \alpha_j {G}_{j,N+1}\label{eq:gamma_1}\\
	\gamma_2&=\Bigg[I_3-\frac{k_0^4}{\varepsilon_0^2}\sum_{i,j=1}^N\alpha_{N+1}  {G}_{N+1,i}  {M}_{ij}^{-1} \alpha_j  {G}_{j,N+1}\Bigg]^{-1}\label{eq:gamma_2}\\
	\gamma_3&=I_3+\frac{k_0^2}{\varepsilon_0}\sum_{i,j=1}^N\alpha_{N+1}  {G}_{N+1,i}  {M}_{ij}^{-1} \frac{ \bm{p}_{0,i} \bm{p}_{0,N+1}^\dag}{ \abs{\bm{p}_{0,N+1}}^2}\label{eq:gamma_3}.
\end{align}
All multiple scattering effects are captured in the three enhancement factors. In general $\gamma_i$ ($i = 1,2,3$) are complex matrices, reflecting the fact that multiple scattering can modify the amplitude, phase and polarization of the scattered field. An estimate of the relative magnitude of the change in the scattered field resulting from multiple scattering $|\delta\bm{E}|/|\delta\bm{E}_{ss}|$ can be found by considering  $\|G(\bm{r},\bm{r}_{N+1}) \gamma_1\gamma_2\gamma_3 G(\bm{r},\bm{r}_{N+1})^{-1} \| \leq \kappa_G \|\gamma_1\gamma_2\gamma_3 \| \leq \kappa_G \|\gamma_1 \|\|\gamma_2 \|\|\gamma_3 \|$ where we have used the sub-multiplicative property of the induced norm and $\kappa_G$ is the condition number of $G(\bm{r},\bm{r}_{N+1})$ {given by the ratio of the maximal and minimal singular values \cite{Golub1996}}. An important class of problems in which equality of the former bound is achieved is systems in which a scalar description is permissible, whereby all tensor quantities ($\alpha_i$, $G$ and $\bm{p}_i$) are replaced with corresponding scalars. In this case, $\kappa_G=1$ and $\abs{\gamma_1\gamma_2\gamma_3}$ directly represents the scaling of the amplitude of $\delta\bm{E}$ from multiple scattering effects.

\begin{figure}[t]
	\centering
	\includegraphics[width=\columnwidth]{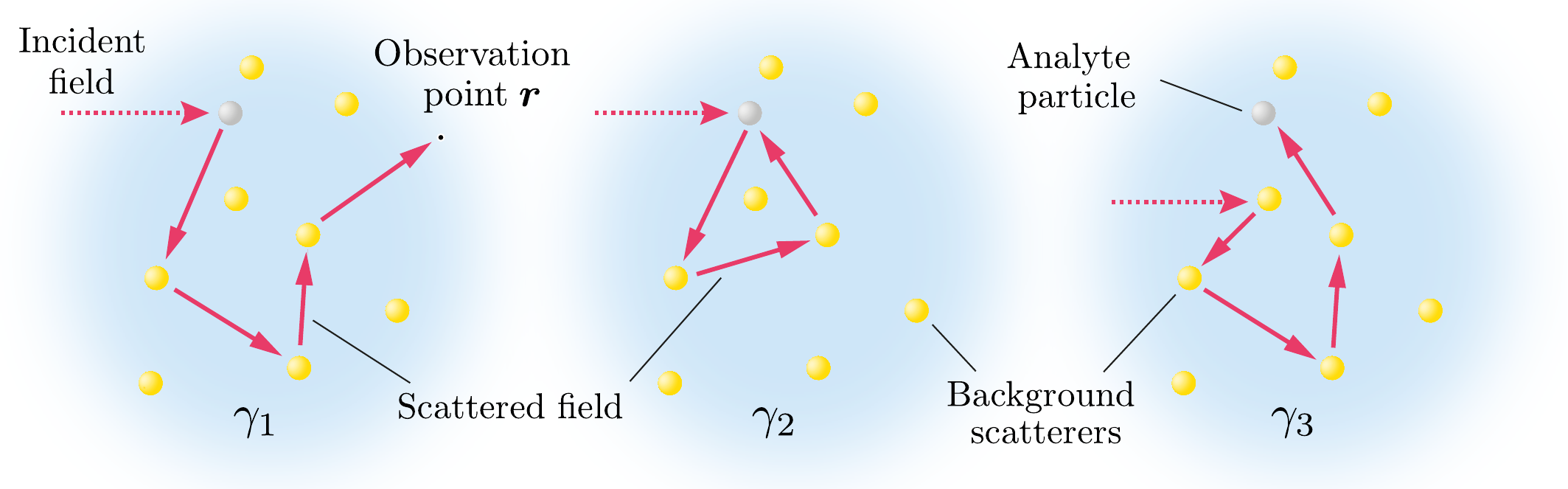}
	\caption{Typical multiple scattering paths associated with each enhancement factor. (left) rescattering of light en-route to the observation point $\bm{r}$ after scattering from the analyte particle, (center) loop trajectories and (right) multiple scattering of the illumination field onto the analyte particle. }
	\label{fig:scattering_paths}
\end{figure}

Physically, each enhancement factor $\gamma_i$ can be associated with a distinct class of multiple scattering trajectories as shown in Figure~\ref{fig:scattering_paths}. Specifically, $\gamma_1$ describes the effect of rescattering of light initially scattered by the analyte particle and hence corresponds to dipole coupling with the analyte. The set of multiple scattering paths described by $\gamma_2$ are closed loops in which light scattered by the analyte particle returns, via scattering off of the initial scatterers, to the analyte particle. Finally, multiple scattering of the incident field onto the added scatterer, which modifies the field at $\bm{r}_{N+1}$, is described by $\gamma_3$. The hotspot effect would manifest in a large value of $\|{\gamma_3}\|$. {The enhancement factors contain a complete description of every possible multiple scattering path. Note that it is possible that $\|\gamma_i\|<1$ and as such the enhancement factors need not describe an increase in the light scattered to a point. Thus, for example, if a particular configuration of scatterers directed light away from the point $\bm{r}$ after leaving the analyte particle, the second term in Eq. \eqref{eq:gamma_1} would (partially) cancel with the first $I_3$ term such that $\delta\bm{E}$ is reduced by the scattering paths described by $\gamma_1$.}

The values of the enhancement factors are determined by the initial scattering configuration ($\alpha_i$ and $\bm{r}_i$ for $i=1,\ldots, N$) and the polarizability and  position of the added analyte particle ($\alpha_{N+1}$ and $\bm{r}_{N+1}$). In reality, however, the exact scattering configuration is rarely known and thus we here study the statistics of the enhancement factors over an ensemble of random configurations. In Ref.~\citenum{Berk2021} we present an analytic treatment of the mean enhancement factors, however, in this letter we consider the average \emph{magnitude} of the enhancement factor, since $\|\gamma_i\|$ is more closely related to experimentally measurable quantities, such as optical intensity. Since a mathematical analysis is not tractable across all scattering regimes, we here use Monte Carlo simulations to study the full range of scatterer densities. For definiteness, we consider multiple scattering of surface plasmon polaritons (SPPs) propagating along a single metal-dielectric interface to illustrate some features of the enhancement factors through evaluation of Eqs.~\eqref{eq:gamma_1}--\eqref{eq:gamma_3}. Notably, SPPs are widely used in biosensors \cite{Homola2003PresentBiosensors} and can play a key role in nanostructured substrates \cite{VanBeijnum2012}. Our example therefore represents an important model system where multiple scattering enhancements can affect single particle sensing and tracking \cite{Zhang2020PlasmonicKinetics,Berk2020TrackingSpeckle}. {A schematic of the system under consideration is shown in the inset of Figure~\ref{fig:gold_dressed_monte_carlo_sim}, in which an SPP propagating along a metal-dielectric interface (with permittivities $\varepsilon_m$ and $\varepsilon_d$ respectively), is scattered from nanoparticles resting on the substrate (see also Ref.~\citenum{Berk2020TrackingSpeckle} for further details).  SPPs can either scatter into other SPPs propagating in a random direction along the metal surface or into waves propagating away from the surface where they are then ultimately detected in the far-field.} {A dipole approximation is valid in this system when surface dressing effects are weak as discussed fully in Ref.~\citenum{Evlyukhin2005Point-dipoleLimitations}}. In such a system our analysis is especially simplified when $|a| \ll 1$, where $a=[\varepsilon_d/(-\varepsilon_m)]^{1/2}$, because a scalar model can be used to describe SPP scattering \cite{Bozhevolnyi1998ElasticExperiment,Evlyukhin2005Point-dipoleLimitations}. Specifically, the relevant scalar field corresponds to the out of plane $E_z$ component of the SPP field, such that only the $G_{zz}$ component of the Green's tensor is considered. The scalar Green's function for points near the surface ($z,z'\ll\lambda_0$) for this model can be approximated as \cite{Evlyukhin2005Point-dipoleLimitations,SoNdergaard,Sndergaard2003VectorialInteractions}
\begin{align}\label{eq:G_SPP}
	G_{\textrm{SPP}}(\bm{r},\bm{r}')&\approx  i A_0e^{-ak_\text{SPP}(z+z')}H_0^{(1)}(k_\text{SPP}\abs{\bm{\rho}-\bm{\rho}'})
\end{align}
where $A_0=ak_\text{SPP} /[2(1-a^4)(1-a^2)]$, $k_\text{SPP}$ is the complex SPP wavenumber (with corresponding absorption length $l_\text{abs}=(2\Im[k_\text{SPP}])^{-1}$) and $H_0^{(1)}(x)$ is the zeroth order Hankel function of the first kind. Eq.~\eqref{eq:G_SPP} is thus used to calculate {$G_{ij}= G_{\text{SPP}}(\bm{r}_i,\bm{r}_j)$}. The elastic SPP scattering cross-section is then given by $\sigma_\text{SPP}=4\abs{\mu}^2/\Re[k_\text{SPP}]$, where $\mu=\alpha(k_0^2/\varepsilon_0) A_0\exp[{-2ak_\text{SPP}z_s}]$ \cite{Bozhevolnyi1998ElasticExperiment}. {Throughout this work we define the elastic scattering mean free path as $l_s=(n\sigma_\text{SPP})^{-1}$, where $n=N/L^2$ is the scatterer density\cite{Akkermans2007MesoscopicPhotons}. Although at large densities the mean free path is more accurately defined in terms of the self energy \cite{Akkermans2007MesoscopicPhotons} we use this parametrisation since the closed form greatly facilitates computation. Deviations from the true length scales are expected, albeit we note all calculations are performed with respect to scatterer density.}

\section*{Results and discussion}
Our Monte-Carlo simulations assumed a free-space wavelength of $\lambda_0=650$~nm, with $\varepsilon_d=1.77$ (corresponding to water) and $\varepsilon_m=-13.68+1.04i$ (corresponding to gold \cite{JohnsonRefractiveIndex}), such that $k_\text{SPP}=(1.42+0.008i)k_0$ {and $|a|=0.36$}. All scatterers were assumed to be identical ($\alpha_i=\alpha~\forall~i$) and  located at a height $z_s$ above the metal interface. Their transverse positions were uniformly randomly distributed on the surface over a square area with sides of length $L$, except for the analyte particle, which was fixed at $\bm{r}_{N+1} = (0,0,z_s)$. The number of scatterers remained fixed at $N=700$, with the scatterer density $n$ adjusted by varying $L$ between $9.4\lambda_0$ and $118\lambda_0$, corresponding to a density ranging from $8\lambda_0^{-2}$ to $0.05\lambda_0^{-2}$. Calculation of the scattered field was performed assuming ${\bm{r}}$ was in the far field. Using a stationary phase approximation to evaluate $G(\bm{r},\bm{r}_i)$  in the far field, reduces these factors, which appear in Eq.~\eqref{eq:gamma_1}, to simple phasors, $G(\bm{r},\bm{r}_{N+1})^{-1}G(\bm{r},\bm{r}_i)=\exp\left[-i\bm{k}_{\text{out}}\cdot(\bm{r}_i-\bm{r}_{N+1})\right]$ where $\bm{k}_{\text{out}}={\varepsilon_d}^{1/2}k_0\hat{\bm{r}}$ is a wavevector in the direction of $\bm{r}$. Specifically, the observation position was taken at $70^\circ$ to the surface normal in the backward $x$ direction ($\bm{k}_{\text{out}}={\varepsilon_d}^{1/2}k_0(-\sin70^\circ,0,\cos70^\circ)$). Results showed only a weak dependence on $\bm{k}_{\text{out}}$. 
The incident field was taken to be a decaying SPP propagating in the $x$ direction of the form $E_{0,z}(x)=\exp(ik_\text{SPP}x)$. With this form, the ratio of dipole moments in Eq. \eqref{eq:gamma_3} reduces to a  form $\sim\exp\left[ik_\text{SPP}(x_i-x_{N+1})\right]$, although since $k_\text{SPP}$ is complex, this factor also describes SPP attenuation. Averages were calculated using 50,000 realisations for each density. {Convergence plots for the worst case scenario are given in the Supporting Information.}

\begin{figure}[t!]
	\includegraphics[width=\columnwidth]{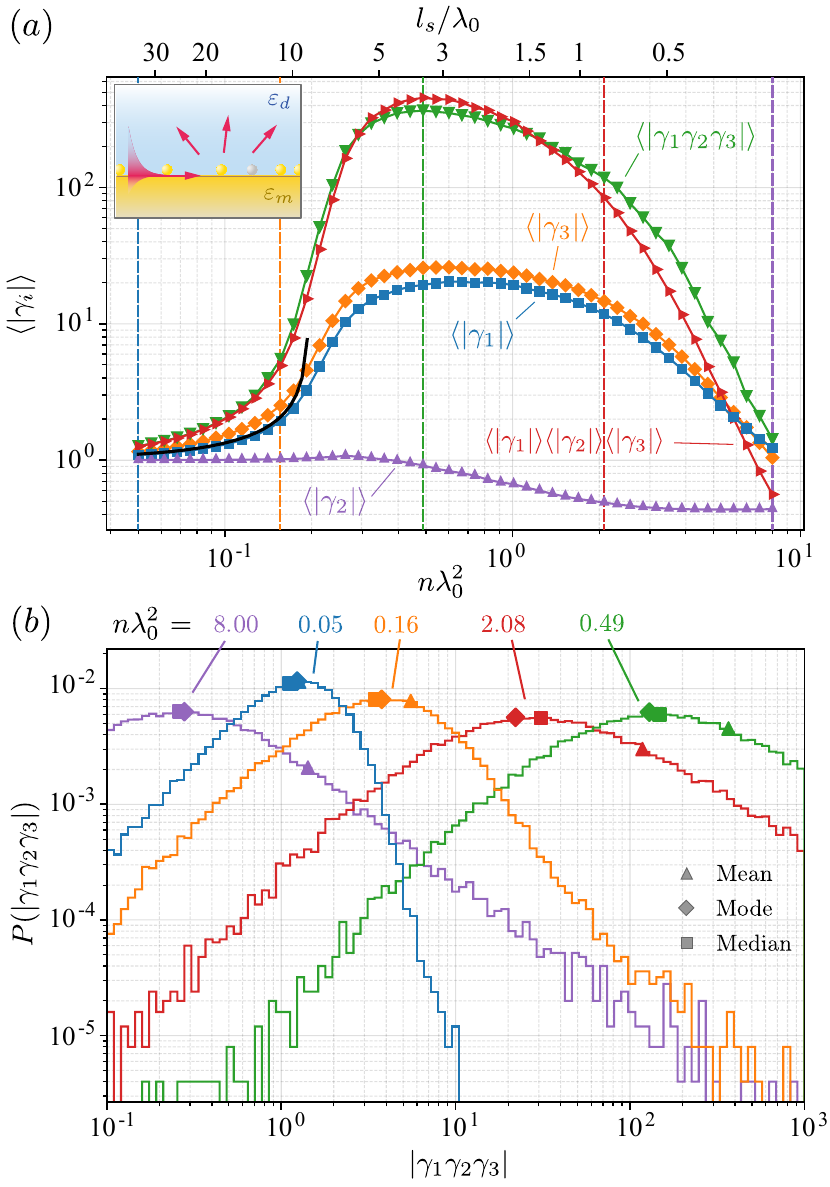}
	\caption{\label{fig:gold_dressed_monte_carlo_sim} Sensing enhancements. (a) Dependence of $\langle\abs{\gamma_1\gamma_2\gamma_3}\rangle$ (green $\triangledown$), $\langle\abs{\gamma_1}\rangle$ (blue $\square$), $\langle\abs{\gamma_2}\rangle$ (purple $\vartriangle$), $\langle\abs{\gamma_3}\rangle$ (orange $\diamond$) and $\langle\abs{\gamma_1}\rangle\langle\abs{\gamma_2}\rangle\langle\abs{\gamma_3}\rangle$ (red $\vartriangleright$) on scatterer density $n$ and mean free path $l_s$ for scatterer polarizability $\alpha = \alpha_g$ corresponding to a 40~nm radius gold nanosphere sitting on the surface ($z_s=40$~nm). The theoretical result from Eq. \eqref{eq:sigma_1} for $\langle\abs{\gamma_1}\rangle$ is also shown (black). {Inset shows schematic of SPP scattering from surface bound nanoparticles.} (b) Relative frequency/probability distributions for the magnitude of the total enhancement $\abs{\gamma_1\gamma_2\gamma_3}$ for scatterer densities of $n\lambda_0^2 = 0.05$ (blue), $0.16$ (orange), $0.49$ (green), $2.08$ (red) and $8.00$ (purple) as also indicated by the corresponding vertical dashed lines in (a). The mean ($\vartriangle$), mode ($\diamond$) and median ($\square$) for each distribution are also shown.}
\end{figure}

The density dependence of the mean total enhancement $\langle\abs{\gamma_1\gamma_2\gamma_3}\rangle$ and the individual mean enhancement factors $\langle\abs{\gamma_i}\rangle$ is shown in Figure~\ref{fig:gold_dressed_monte_carlo_sim}(a) for particle polarizability, $\alpha_g=(3.74 + 0.33i)\times10^{-32}\textrm{~Cm}^2\textrm{V}^{-1}$, corresponding to a dressed 40~nm radius gold sphere sat on the gold film ($z_s=40$~nm). For the given parameters, the density range simulated corresponds to a scattering mean free path varying from $34.3\lambda_0$ down to $0.21\lambda_0$. The mean enhancement factor initially increases with density and rises above 1, indicating that multiple scattering on average enhances the sensitivity at these lower densities. As scatterer density increases further the mean enhancement reaches a maximum of $\sim 367$ at an optimal density of $n=0.49 / \lambda_0^{2}$ ($l_s = 3.51 \lambda_0$), before then decreasing at higher $n$, eventually dropping below one, indicating that at extremely high densities, multiple scattering acts to decrease the scattered signal perturbation on average. {We attribute this decrease to SPP localisation \cite{Zhu2020spp} effects which restrict the impact of the additional particle to a region of the order of the localization length in size. In particular, we note the localisation length of a 2D system can be estimated as $\xi=l_s\exp\left(\pi \text{Re}[k_{\text{SPP}}]l_s/2\right)$ \cite{Sheng1995}, which becomes comparable to the system size for $l_s \approx 0.73\lambda_0$ in our simulations. Specifically, for $l_s = 0.21 \lambda_0$ we have $\xi/L = 0.42$. Note it has been shown that Anderson localisation of light cannot be achieved for fully vectorial 3D random ensembles of dipole scatterers, such that we would not expect a corresponding decrease in enhancement for such systems \cite{Skipetrov2014}.} In general, $\langle\abs{\gamma_2}\rangle$ remains close to one, meaning the effect of loop paths is weak compared to $\langle\abs{\gamma_1}\rangle$ and $\langle\abs{\gamma_3}\rangle$ which are of comparable magnitude.

An approximate scaling theory for the behaviour of $\langle |\gamma_{1,3}|\rangle$ in the low density regime can be derived by treating the sums in Eqs.~\eqref{eq:gamma_1} and \eqref{eq:gamma_3} as random phasor sums. Specifically, when $l_s$ is larger than $\lambda_\text{SPP}=2\pi/\Re[k_\text{SPP}]$, propagation between each scattering event decorrelates the amplitude and phase of each phasor in the sum such that the sums are circular Gaussian random variables with variance $\sigma_{1,3}^2=N\langle\abs{A_i}^2\rangle/2$ where $\abs{A_i}$ are the amplitudes of the elements of the corresponding sum \cite{goodman2007speckle}. The amplitude of $\gamma_{1,3}$ thus follows a Rician distribution with width parameter $\sigma_{1,3}$. For $\gamma_1$ we have $A_i=({k_0^2}/{\varepsilon_0})e^{-i\bm{k}_{\text{out}}\cdot(\bm{r}_i-\bm{r}_{N+1})}\sum_jM_{ij}^{-1}\alpha_jG_{j,N+1}$, which represents the sum of all scattering paths from $\bm{r}_{N+1}$ to $\bm{r}_i$. In calculating $\langle\abs{A_i}^2\rangle$, the interference of all paths should be considered however, adopting the ladder approximation (valid when $k_\text{SPP}l_s\gg1$), only the interference of identical scattering paths are assumed to contribute to the average owing to the random phase difference between different trajectories \cite{Akkermans2007MesoscopicPhotons}. Within this approximation, we find in the limit $N,L\to\infty$ with $n=N/L^2$ fixed (see Supporting Information)
\begin{equation}\label{eq:sigma_1}
	\sigma_1^2=\frac{1}{2}\frac{l_s^{-1}}{l_\text{abs}^{-1}+4n\Im(\mu)/(\Re[k_\text{SPP}])-l_s^{-1}}.
\end{equation}
In the lossless case ($\Im[k_\text{SPP}]=0$) $\sigma_3^2$ is identical to $\sigma_1^2$. Using the properties of the Rician distribution, the resulting mean magnitude of the enhancement follows as $\langle\abs{\gamma_{1,3}}\rangle=\sigma_{1,3}(\pi/2)^{1/2}L_{1/2}(-1/(2\sigma_{1,3}^2))$, where $L_{1/2}(x)$ is a generalized Laguerre polynomial. Since $l_s^{-1}$ is proportional to $n$, $\langle\abs{\gamma_{1}}\rangle$ initially increases from 1 linearly with density, before increasing much more rapidly as $l_s$ approaches $l_\text{abs}$. The result diverges when the denominator vanishes, by which point the ladder approximation breaks down and the effects of interference between different paths (such as coherent backscattering) become significant \cite{Akkermans2007MesoscopicPhotons}. This behaviour is  evidenced in Figure~\ref{fig:gold_dressed_monte_carlo_sim} with good agreement found between the ladder approximation for $\langle\abs{\gamma_1}\rangle$ and numerical calculations over the range of validity. The density dependence of $\langle\abs{\gamma_3}\rangle$ is analogous, however the effect of loss (included in the numerical simulations) is to slightly increase $\langle\abs{\gamma_3}\rangle$. 

In general, the individual enhancement factors are not statistically independent such that $\langle\abs{\gamma_1\gamma_2\gamma_3}\rangle \neq \langle\abs{\gamma_1}\rangle\langle\abs{\gamma_2}\rangle\langle\abs{\gamma_3}\rangle$, as also shown in Figure~\ref{fig:gold_dressed_monte_carlo_sim}(a). Qualitative agreement between $\langle\abs{\gamma_1\gamma_2\gamma_3}\rangle$ and $\langle\abs{\gamma_1}\rangle\langle\abs{\gamma_2}\rangle\langle\abs{\gamma_3}\rangle$ is clearly apparent, particularly at lower densities, however correlations cause a noticeable quantitative difference at densities at or beyond the peak. Analysis of the Pearson's correlation coefficients $P_{ij}$ between $\abs{\gamma_i}$ and $\abs{\gamma_j}$ ($i\neq j)$, reveals that $|\gamma_2|$ shows little correlation with the other enhancement factors ($P_{12}, P_{23}\in [-0.1,0.1]$) across the full density range. This is because the loop paths associated with $\gamma_2$ are distinct from the scattering paths in $\gamma_{1,3}$. In contrast, scattering trajectories contributing to $\gamma_1$ and $\gamma_3$ are partially related by reciprocity \cite{Byrnes2021}, such that a multiple scattering path from $\bm{r}_{N+1}$ to $\bm{r}_i$ (associated with $\gamma_1$) has the same phase and amplitude as the reciprocal path going from $\bm{r}_i$ to $\bm{r}_{N+1}$ (associated with $\gamma_3$). Correlation of $\abs{\gamma_1}$ and $\abs{\gamma_3}$ is hence dictated by the correlation between the additional propagation phases appearing in each enhancement factor, namely that of propagation of the scattered (incident) field from (to) the relevant scattering particle. At low densities, these propagation phases remain uncorrelated ($\abs{P_{13}} \lesssim 0.1$ for $n\lambda_0^2 \lesssim 0.1$), however, at higher densities the typically shorter distances between scattering sites and the analyte particle mean the phase difference of the incident and outgoing fields are smaller resulting in increased correlation ($P_{13} \in [0.6,0.8]$ for $n\lambda_0^2>0.2$).

Histograms of the relative frequency of $\abs{\gamma_1\gamma_2\gamma_3}$, shown in Figure~\ref{fig:gold_dressed_monte_carlo_sim}(b), demonstrate that at  low densities the distribution of total enhancements is tightly centred around $\sim 1$. At densities close to the optimum value, the probability distribution however exhibits a long tail. A given scattering configuration at the optimum density consequently has a high probability of producing a significant sensitivity enhancement, however it should be noted that the total enhancement for a given realisation will likely be smaller than the mean total enhancement (mode $\approx$ median $\ll$ mean), typically $\sim 100$. Importantly, there is a small but non-negligible probability of a very large enhancement even  as high as $\sim 10^3$. At the highest densities, the majority of realisations suppress sensitivity, albeit the tail is still longer relative to the lowest densities. Consequently, even though the mean enhancements for the two limiting cases are both of order unity, for high scatterer density there exist a small number of configurations that produce an appreciable sensitivity enhancement. In contrast, at low densities, different configurations do not differ greatly in their effect on sensitivity.

\begin{figure}[t]
	\includegraphics[width=\columnwidth]{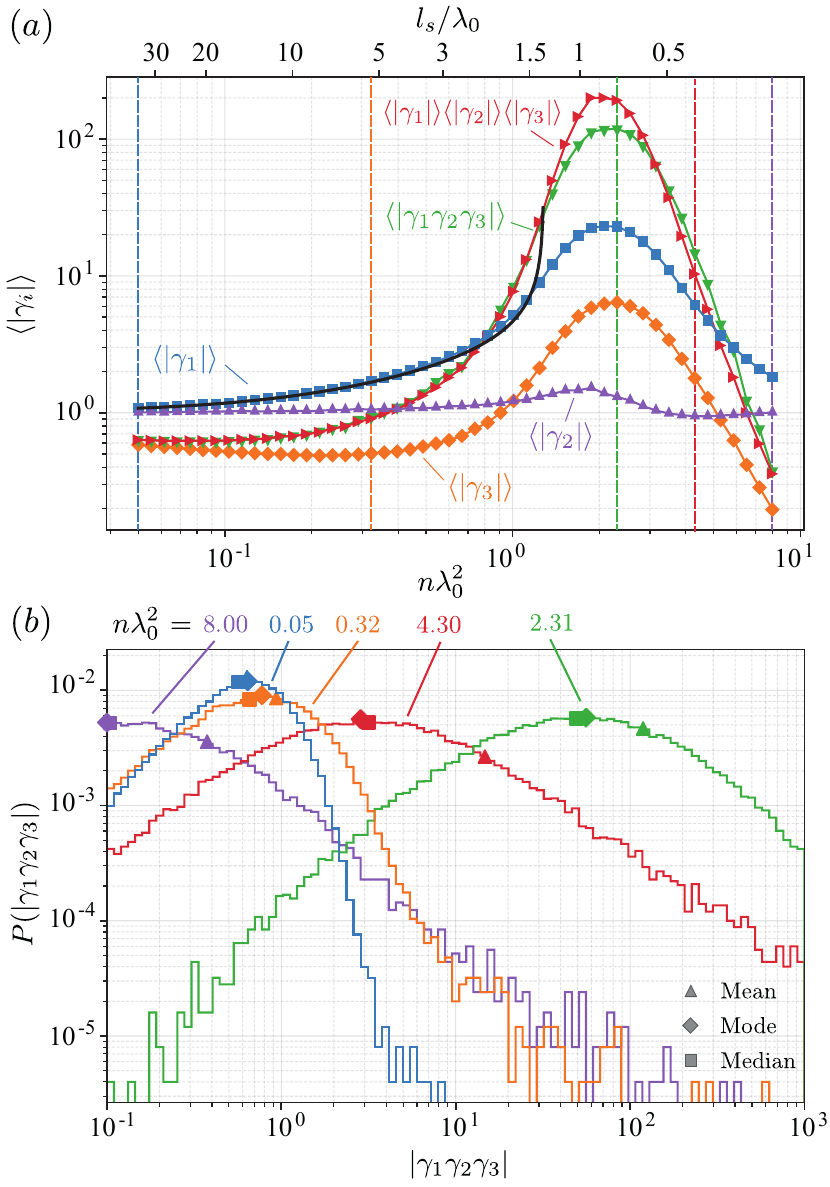}%
	\caption{Sensing enhancements for phase-shifted polarizability. As Figure~\ref{fig:gold_dressed_monte_carlo_sim} albeit for polarizability $\alpha = \alpha_g e^{i \pi/2}$.}
	\label{fig:phase_shifted_monte_carlo_sim}
\end{figure}

Importantly, Eqs.~\eqref{eq:gamma_1}--\eqref{eq:gamma_3}, predict that the statistics of the total enhancement are sensitive to the phase of $\mu$ by virtue of the  $\alpha G_{ij}$ factors. Physically, this parameter can be tuned in multiple ways. Variation of either the material composition or geometrical properties of the individual scatterers can, for example, modify the particle polarizability $\alpha$. Moreover, for resonant scatterers, such as plasmonic nanoparticles, tuning the operational wavelength provides an additional degree of freedom. Introduction of an index matched spacer layer between the substrate and background scatterers furthermore allows the height $z_s$ to be adjusted. Shifting the phase of $\mu$ whilst holding its amplitude constant leaves both the elastic SPP scattering cross-section and mean free path unchanged, however results in a change in the absorption cross-section and/or scattering out of SPPs. Consequently, a different dependence of the mean enhancement on scatterer density is seen as shown in Figure \ref{fig:phase_shifted_monte_carlo_sim} for a phase shift of $\pi/2$. Notably, in this case the mean enhancements are reduced at low densities compared to Figure \ref{fig:gold_dressed_monte_carlo_sim}, which we attribute to a reduction in the field incident on scatterers due to increased absorption and scattering out of SPP modes. At higher $n$, however, the same decay in enhancement with increasing density is seen. The maximum sensitivity enhancement is of similar magnitude ($\sim 119$) and occurs at a higher density ($n=2.31/\lambda_0^2$, $l_s=0.74\lambda_0$) compared to the gold nanosphere case. Enhancements are furthermore seen to occur over a narrower density range. Good agreement between the ladder approximation for $\langle\abs{\gamma_1}\rangle$ is once more evident, however, $\langle\abs{\gamma_3}\rangle$ is significantly reduced, due to the increased role played by absorption. The probability distributions shown in Figure \ref{fig:phase_shifted_monte_carlo_sim}(b) show the same behaviour as the gold sphere case in the low density, near-optimal density and high density regimes, however the transition between each regime occurs at different densities. Similarly,  $P_{ij}$ show similar trends as for the gold, although correlations between $|\gamma_1|$ and $|\gamma_3|$ become noticeable at a higher density.

\section*{Conclusions}
Using a coupled dipole formalism, we have derived expressions for the multiple scattering based enhancement of the scattered field perturbation when a scatterer is added to a disordered scattering environment. Eqs.~\eqref{eq:gamma_1}--\eqref{eq:gamma_3} apply quite generally to a range of wave scattering phenomena, both vector and scalar, through appropriate choice of the Green's tensor. The total enhancement factor derives from three contributions, each arising from different sets of multiple scattering paths, hence allowing insight into the physical mechanisms that affect single particle sensitivity in the multiple scattering regime. {Although the  local density of states (LDOS) \cite{Carminati2015} is frequently used to assess the effect of spatial inhomogeneity, such as system disorder, on an oscillating dipole, it is important to note that the enhancement factors introduced here capture important additional features present in the sensing system considered. The LDOS describes the relative power radiated by a dipole in an inhomogeneous environment compared to free space and would hence describe the enhancement for e.g. dark-field scattering based  or fluorescence detection \cite{Weigel2014}, however, in the system considered in this work particle detection exploits interferometric detection  \cite{Taylor2019InterferometricScattering,Ignatovich2006}. Specifically, the illumination field generates a background field which coherently interferes with the field scattered from an analyte particle. Accordingly, the magnitude of the scattered signal scales as $R^3$ as opposed to $R^6$, where $R$ is the analyte particle radius, hence crucially helping to mitigate noise. Nevertheless, both the LDOS and the enhancement factors of Eqs.~\eqref{eq:gamma_1}--\eqref{eq:gamma_3} derive from the system Green's function and similar features, such as an exponential distance dependence \cite{Castanie2012} and long tailed decay \cite{Krachmalnicoff2010,Riboli2017}, are seen.}

{Based on our model,} Monte Carlo simulations of SPP scattering by dipole scatterers randomly distributed on a metal-dielectric interface were performed, which demonstrated that the sensitivity to addition of a single particle can be enhanced by a factor of order $10^2$ on average. Moreover, it was shown that there exists an optimum density of scatterers at which the sensitivity gain is maximised. While the optimum density depends on the properties of the individual scatterers, the size of the peak enhancement is relatively insensitive to the individual scatterers. Our results can hence be used to optimise the design of SPP sensors consisting of random nano-scatterers in order to maximise sensitivity.
Physically, the optimum scatterer density exists due to the competing effects of dipole coupling and Anderson localisation and would thus be expected in a range of disordered systems beyond the SPP scattering considered in this work. Whilst the former effect typically increases the average scattering perturbation induced by addition of an analyte particle, the latter confines such perturbations to a smaller spatial region. Optimal configurations would however not be expected in scattering systems in which localisation is more difficult or cannot be achieved, such as 3D electromagnetic scattering in ensembles of point scatterers \cite{Skipetrov2014}.

Finally, we note that fabrication of nanostructures and deposition of nanoparticles of sub-wavelength dimensions is becoming more routine, using methods such as electron beam lithography and focused ion beam lithography \cite{Stewart2008}, such that the dipole approximation made in this work is applicable to realistic experimental systems. Moving beyond dipole scatterers to larger structures does however introduce preferential  scattering in the forward direction. Similar anisotropic scattering can also occur for SPP scattering when surface dressing is large \cite{Evlyukhin2005Point-dipoleLimitations}. In such scenarios, the transport mean free path $l_{tr} = l_s /(1+\langle \cos \theta \rangle)$, where $\langle \cos \theta \rangle$ is the average of cosine of the scattering angle \cite{Akkermans2007MesoscopicPhotons}, describes the length scale over which the scattering direction is randomised and therefore represents a more suitable parameterisation of different scattering regimes. For highly anisotropic scattering the transport mean free path can however become very long, such that densities required to achieve localisation are difficult to reach. Moreover, in systems with loss such as SPP sensors, the absorption length must be longer than the mean free path for multiple scattering effects and localisation to play a role. Such factors must therefore also be considered when optimising sensitivity of random nanostructured sensors \cite{Berk2021}.

	\begin{acknowledgments}
		This work was funded by the Engineering and Physical Sciences Research
		Council (EPSRC) (1992728) and the Royal Society (UF150335).
	\end{acknowledgments}

\end{document}